\documentclass[aps, prl, amsmath, superscriptaddress, twocolumn, nofootinbib]{revtex4-1}

\usepackage{amsmath}
\usepackage{txfonts}
\usepackage{graphicx}
\usepackage{threeparttable}
\usepackage{latexsym}
\usepackage{amsbsy}
\usepackage{float}
\usepackage{mathrsfs}
\usepackage{color}
\usepackage{color,bm}
\usepackage{array}
\newcommand{\PreserveBackslash}[1]{\let\temp=\\#1\let\\=\temp}
\newcolumntype{C}[1]{>{\PreserveBackslash\centering}p{#1}}

\makeatletter
\def\ignorespacesandimplicitepars{%
  \begingroup
  \catcode13=10
  \@ifnextchar\relax
    {\endgroup}%
    {\endgroup}%
}

\renewcommand{\section}[1]{\emph{#1.}---\ignorespacesandimplicitepars}
\renewcommand{\subsection}[1]{}

\let\jnl@style=\rm
\def\ref@jnl#1{{\jnl@style#1}}

\def\aj{\ref@jnl{Astron.~J.}}                   
\def\araa{\ref@jnl{ARA\&A}}             
\def\apj{\ref@jnl{Astrophys.~J.}}                 
\def\apjl{\ref@jnl{ApJ}}                
\def\apjs{\ref@jnl{Astrophys.~J.~Suppl.~Ser.}}               
\def\ao{\ref@jnl{Appl.~Opt.}}           
\def\apss{\ref@jnl{Ap\&SS}}             
\def\aap{\ref@jnl{Astron.~Astrophys.}}                
\def\aapr{\ref@jnl{A\&A~Rev.}}          
\def\aaps{\ref@jnl{A\&AS}}              
\def\azh{\ref@jnl{AZh}}                 
\def\baas{\ref@jnl{BAAS}}               
\def\jrasc{\ref@jnl{JRASC}}             
\def\memras{\ref@jnl{MmRAS}}            
\def\mnras{\ref@jnl{Mon.~Not.~R.~Astron.~Soc.}}             
\def\pra{\ref@jnl{Phys.~Rev.~A}}        
\def\prb{\ref@jnl{Phys.~Rev.~B}}        
\def\prc{\ref@jnl{Phys.~Rev.~C}}        
\def\prd{\ref@jnl{Phys.~Rev.~D}}        
\def\pre{\ref@jnl{Phys.~Rev.~E}}        
\def\prl{\ref@jnl{Phys.~Rev.~Lett.}}    
\def\pasp{\ref@jnl{PASP}}               
\def\pasj{\ref@jnl{Publ.~Astron.~Soc.~Jpn}}               
\def\qjras{\ref@jnl{QJRAS}}             
\def\skytel{\ref@jnl{S\&T}}             
\def\solphys{\ref@jnl{Sol.~Phys.}}      
\def\sovast{\ref@jnl{Soviet~Ast.}}      
\def\ssr{\ref@jnl{Space~Sci.~Rev.}}     
\def\zap{\ref@jnl{ZAp}}                 
\def\nat{\ref@jnl{Nature}}              
\def\iaucirc{\ref@jnl{IAU~Circ.}}       
\def\aplett{\ref@jnl{Astrophys.~Lett.}} 
\def\apspr{\ref@jnl{Astrophys.~Space~Phys.~Res.}}
\def\bain{\ref@jnl{Bull.~Astron.~Inst.~Netherlands}} 
\def\fcp{\ref@jnl{Fund.~Cosmic~Phys.}}  
\def\gca{\ref@jnl{Geochim.~Cosmochim.~Acta}}   
\def\grl{\ref@jnl{Geophys.~Res.~Lett.}} 
\def\jcp{\ref@jnl{J.~Chem.~Phys.}}      
\def\jgr{\ref@jnl{J.~Geophys.~Res.}}    
\def\jqsrt{\ref@jnl{J.~Quant.~Spec.~Radiat.~Transf.}}
\def\memsai{\ref@jnl{Mem.~Soc.~Astron.~Italiana}}
\def\nphysa{\ref@jnl{Nucl.~Phys.~A}}   
\def\physrep{\ref@jnl{Phys.~Rep.}}   
\def\physscr{\ref@jnl{Phys.~Scr}}   
\def\planss{\ref@jnl{Planet.~Space~Sci.}}   
\def\procspie{\ref@jnl{Proc.~SPIE}}   
\def\rmp{\ref@jnl{Rev.~Mod.~Phys.}} 
\def\grg{\ref@jnl{Gen.~Relativ.~Gravit.}} 
\def\ap{\ref@jnl{Ann.~Phys.}} 
\def\lrr{\ref@jnl{Living~Rev.~Relat.}} 
\def\jcap{\ref@jnl{J.~Cosmol.~Astropart.~Phys.}} 
\def\raa{\ref@jnl{Res.~Astron.~Astrophys.}} 

\makeatother

\begin{document}

\preprint{}

\title{Constraining $f(R)$ Gravity Theory Using Weak Lensing Peak Statistics from the Canada-France-Hawaii-Telescope Lensing Survey}

\author{Xiangkun Liu}
\email[]{lxk98479@pku.edu.cn}
\affiliation{Department of Astronomy, School of Physics, Peking University, Beijing 100871, People's Republic of China}
\author{Baojiu Li}
\affiliation{Institute for Computational Cosmology, Department of Physics, Durham University, South Road, Durham DH1 3LE, United Kingdom}
\author{Gong-Bo Zhao}
\affiliation{National Astronomy Observatories, Chinese Academy of Sciences, Beijing 100012, People's Republic of China}
\affiliation{Institute of Cosmology \& Gravitation, University of Portsmouth, Dennis Sciama Building, Portsmouth PO1 3FX, United Kingdom}
\author{Mu-Chen Chiu}
\affiliation{The Shanghai Key Lab for Astrophysics, Shanghai Normal University, 100 Guilin Road, Shanghai 200234, People's Republic of China}
\author{Wei Fang}
\affiliation{The Shanghai Key Lab for Astrophysics, Shanghai Normal University, 100 Guilin Road, Shanghai 200234, People's Republic of China}
\affiliation{Department of Physics, Shanghai Normal University, 100 Guilin Road, Shanghai 200234, People's Republic of China}
\author{Chuzhong Pan}
\affiliation{Department of Astronomy, School of Physics, Peking University, Beijing 100871, People's Republic of China}
\author{Qiao Wang}
\affiliation{Key Laboratory for Computational Astrophysics, The Partner Group of Max Planck Institute for Astrophysics, National Astronomical Observatories,
Chinese Academy of Sciences, Beijing 100012, People's Republic of China}
\author{Wei Du}
\affiliation{National Astronomy Observatories, Chinese Academy of Sciences, Beijing 100012, People's Republic of China}
\author{Shuo Yuan}
\affiliation{Department of Astronomy, School of Physics, Peking University, Beijing 100871, People's Republic of China}
\author{Liping Fu}
\affiliation{The Shanghai Key Lab for Astrophysics, Shanghai Normal University, 100 Guilin Road, Shanghai 200234, People's Republic of China}
\author{Zuhui Fan}
\affiliation{Department of Astronomy, School of Physics, Peking University, Beijing 100871, People's Republic of China}
\affiliation{Collaborative Innovation Center of Modern Astronomy and Space Exploration, Nanjing 210093, People's Republic of China}

\date{\today}

\begin{abstract}

In this Letter, we report the observational constraints on the Hu-Sawicki $f(R)$ theory derived from weak lensing peak abundances,
which are closely related to the mass function of massive halos. In comparison with studies using optical or x-ray clusters of galaxies,
weak lensing peak analyses have the advantages of not relying on mass-baryonic observable calibrations. 
With observations from the Canada-France-Hawaii-Telescope Lensing Survey, our peak analyses give rise to a tight constraint on 
the model parameter $|f_{R0}|$ for $n=1$. The $95\%$ C.L. is $\log_{10}|f_{R0}| < -4.82$ given WMAP9 priors on $(\Omega_{\rm m}, A_{\rm s})$.
With Planck15 priors, the corresponding result is $\log_{10}|f_{R0}| < -5.16$.    

\end{abstract}


\maketitle

\section{Introduction}

While both are able to explain the observed late-time accelerating expansion of the universe \citep{Riess1998, Perlmutter1999}, 
modified gravity (MG) theories \citep[e.g.,][]{Koyama2008, DurMaar2008, JaKh2010} and dark energy models 
in general relativity (GR)
\citep[e.g.,][]{Weinberg1989} lead to different formation and evolution of cosmic structures 
\citep[e.g.,][]{Schmidt2009, Zhao2011a, Zhao2011b, Zhao2012, Li2012a, Li2012b, Li2013, Zhao2014, Achitouv2015,Joyce2015,Chiu2015,Stark2016}. 
Observations of large-scale structures are therefore critical in scrutinizing the underlying mechanism driving 
the global evolution of the universe and in revealing the fundamental law of gravity. 

The $f(R)$ theory is a representative MG model, 
in which the integrand of the Einstein-Hilbert action is  
$R+f(R)$, where $f(R)$ is a function of the scalar curvature $R$ \citep{Sotiriou2010, deFelice2010, Zhao2011a}.
By choosing $f(R)$ properly, such as the Hu-Sawicki model \citep[][hereafter HS07]{HuSaw2007}, 
the theory can give rise to the late time cosmic acceleration without violating 
the gravity tests in the solar system and without affecting high redshift physics significantly.  
Matching the expansion history with that of the flat $\Lambda$CDM model with the matter density parameter $\Omega_{\rm m}$,
an extra degree of freedom is $f_R=df/dR$.
For HS07, $f_R\approx -n(c_1/c_2^2)[m^2/(-R)]^{n+1}$ (with the sign convention used in \citet{Zhao2011a}),
and its current background value is $f_{R0}\approx -n(c_1/c_2^2)[3(1+4\Omega_{\Lambda}/\Omega_{\rm m})]^{-(n+1)}$.
Here $m^2=H_0^2\Omega_{\rm m}$ with $H_0$ being the present Hubble constant, 
$c_1/c_2=6\Omega_\Lambda/\Omega_{\rm m}$, and $\Omega_\Lambda=1-\Omega_{\rm m}$.
It satisfies the solar system tests for $n\ge 1$
\citep[HS07,][]{Zhao2011a}. On the other hand, cosmic structures can be affected significantly.
Thus independent observational studies of different scales are important in probing the nature of gravity
\citep[e.g.,][]{Jain2013s, Weinberg2013, Jain2013, Higuchi2016}.

On cosmological scales, there have been different observational analyses \citep[e.g.,][]{Oyaizu2008, Raveri2014, Planck15XIV}. 
Among them, studies of clusters of galaxies provide the most sensitive constraints 
\citep[e.g.,][]{Schmidt2009, Ferraro2011, Lombriser2012a, Cataneo2015, Wilcox2015, Li2016} and reach a level 
 of $\log_{10}|f_{R0}|< -4.8 $ ($95\%$ C.L.) \citep{Cataneo2015}.  

Weak lensing effects (WL) are a key cosmological probe
\citep[e.g.,][]{BartSch2001, Albrecht2006, LSST2012, Amendola2013, Weinberg2013, FuFan2014}.
Cosmic shear correlation analyses have been incorporated to constrain gravity theories \citep[e.g.,][]{Harnoise2015}.
WL peak statistics, particularly high peaks, possess the cosmological sensitivities of both WL effects and 
massive clusters, and provide an important complement to shear correlation studies
\citep[e.g.,][]{White2002,Hamana2004, TangFan2005, DietHart2010, Fan2010, Yang2011, Hamana2012, Liu2014, Lin2015, Liujia2015, Liu2015}. 
In comparison with cluster studies that normally involve baryonic observables,
WL peak analyses are advantageous because of the gravitational origin of WL effects.  

In this Letter, we derive constraints on the HS07 model parameter $|f_{R0}|$ for $n=1$, for the first time, from WL peak abundances using WL data from Canada-France-Hawaii-Telescope Lensing Survey (CFHTLenS) \citep{Erben2013}.
We perform mock tests to validate our pipeline before applying to actual data analyses.

\section{Observational data}

CFHTLenS covers a total survey area of $\sim 154\deg^2$ from
171 individual pointings distributed in four regions \citep{Erben2013}. We note that for cosmic shear correlation analyses, 
129 pointings pass the systematic tests \citep{Heymans2012}. For the high peak abundances, our analyses find 
that using the full pointings does not introduce any notable bias comparing to that using the passed fields. 
We therefore keep the full data here. The photometric redshift is estimated for each galaxy from five bands $u^*g'r'i'z'$ observations 
\citep{Hildebrandt2012}. The forward modelling LENSFIT pipeline is applied for the shape measurement \citep{Miller2013}. 
After masking out bright stars and faulty CCD rows across the entire survey, the effective survey area is $\sim 127\deg^2$. 
We select source galaxies with weight $w>0$, $\hbox{FITCLASS}=0$, $\hbox{MASK}\le 1$ and redshift in the range $z=[0.2, 1.3]$ 
in our weak lensing analyses \citep{Miller2013}. 
Such a selection results in a total number of 5,596,690 source galaxies.
By taking into account their weights, the effective number of galaxies is $\sim 4.5 \times 10^6$, corresponding to
the density $\sim 10 \hbox{ arcmin}^{-2}$. By summing up the photo-z probability distribution of each source galaxy, 
we obtain the redshift distribution for our source sample with $p_z(z)=A{(z^a+z^b)}/{(z^c+d)}$ and $A=0.5514$, $a=0.7381$, $b=0.7403$, 
$c=6.0220$ and $d=0.6426$.

\section{Peak analyses}

We perform WL peak analyses following the procedures described in detail in \citet{Liu2015}.
The steps are briefly summarized here. (1) We calculate the smoothed shear field taking into account properly 
the additive and multiplicative bias corrections. Then the convergence $\kappa$ map is reconstructed for each individual $\sim 1\times 1\deg^2$ field using the nonlinear Kaiser-Squires method \citep[e.g.,][]{KS1993, Bartel1995, SK1996}. The corresponding smoothed filling-factor map is also generated from the positions and weights of source galaxies. We apply
a Gaussian smoothing with $W_{\theta_{\rm G}}(\boldsymbol{\theta})=1/({\pi\theta_{\rm G}^2})
\exp{(-{|\boldsymbol{\theta}|^2}/{\theta_{\rm G}^2})}$ taking $\theta_{\rm G}=1.5$ arcmin.  
(2) For each convergence map defined on $1024\times 1024$ pixels, we identify peaks by comparing their $\kappa$ values 
with those of their nearest $8$ neighboring pixels. We exclude regions with the filling factor values $\le 0.5$ in peak counting to suppress the mask effects \citep{Liu2014,Liu2015}, and also the outer most $50$ pixels
in each side of an individual map to eliminate the boundary effect. The total leftover area for peak counting is $\sim 112\hbox{ deg}^2$. (3) We divide peaks into different bins based on their signal to noise ratio $\nu=\kappa/\sigma_0$, where $\sigma_0$ is the average rms of the shape noise estimated by randomly rotating source galaxies to construct noise maps. For CFHTLenS and with $\theta_{\rm G}=1.5$ arcmin, $\sigma_0\approx 0.026$. 
In this paper, we only consider high peaks with $\nu\ge 3$. To avoid possible bias arising from a single bin with very few peaks 
and thus a large statistical fluctuation, we adopt unequal binning with comparable numbers of peaks in different bins, specifically
$\nu=[3, 3.1], (3.1, 3.25], (3.25, 3.5], (3.5, 4], (4, 6]$. The peak counts are then denoted by $N_i^d$ $(i=1,...,5)$. 

To derive cosmological constraints from peak counts, we define the following $\chi^2$ to be minimized \citep{Liu2015}
\begin{equation}
\chi_{\rm peak}^{2}=\boldsymbol{dN}^{(p')}(\widehat{\boldsymbol{C}^{-1}})\boldsymbol{dN}^{(p')},
\label{chipeak}
\end{equation}
where $\boldsymbol{dN}^{(p')}=\boldsymbol{N}^d-\boldsymbol{N}^{p'}$ is the difference between the data vector 
$\boldsymbol {N}^d$ and the theoretical expectations of the peak counts $\boldsymbol {N}^{p'}$ for the cosmological model $p'$.
The covariance matrix $\boldsymbol{C}$ is estimated from bootstrap analyses using the CFHTLenS data themselves. 
The matrix $\widehat{\boldsymbol{C}^{-1}}$ is the scaled inverse covariance matrix with 
$\widehat{\boldsymbol{C}^{-1}}=(R_{\rm s}-N_{\rm bin}-2)/(R_{\rm s}-1)(\boldsymbol{C}^{-1})$, where
$N_{\rm bin}=5$, and $R_{\rm s}=10,000$ is the total number of bootstrap samples.

For $\boldsymbol {N}^{p'}$, we use the theoretical model
of \citet[][here after F10]{Fan2010}. The model assumes that a true high peak is contributed dominantly from a single massive halo. 
The shape noise effects, the major contaminations to WL peak analyses using relatively shallow surveys, such as
CFHTLenS, are fully accounted for.
The cosmological quantities involved in F10 are the mass function and the internal density profile of dark matter halos, and the
cosmological distances in the lensing efficiency factor as well as in the volume element.  
F10 has been tested extensively by comparing with simulations \citep{Liu2014, Liu2015, Lin2015}.
It has also been applied to derive cosmological constraints, within the framework of $\Lambda$CDM model, from observed WL peaks \citep{Liu2015}. 

We adopt the halo mass function given by \citet{Kopp2013}, valid for $10^{-7}\le |f_{R0}|\le 10^{-4}$.
We compare its predictions with that from our $f(R)$ simulations to be described in the 
next section, and find a good agreement.
For the halo density profile in $f(R)$ theory, studies have shown
that it is not different significantly from that of the corresponding $\Lambda$CDM model for massive halos
concerned in our peak analyses here \citep[e.g.,][]{Zhao2011a, Lombriser2012b, Achitouv2015, Shi2015}.
We therefore use the Navarro-Frenk-White (NFW) density profile \citep{NFW1996, NFW1997} with the mass-concentration (M-c) relation given by \citet{Duffy2008}. 
We have checked and found that different choices of M-c relation do not affect our constraint on $|f_{R0}|$ significantly
due to the weak degeneracy between these two parameters from current data.
We should note that in both the mass function of \citet{Kopp2013} and in our $f(R)$ simulations and observational analyses, the $\sigma_8$ parameter, the 
rms of the present linearly extrapolated density perturbations smoothed with a top-hat window function of scale $8h^{-1}\hbox{Mpc}$, 
is defined to be the $\Lambda$CDM-equivalent value rather than its true value in $f(R)$ theory. Thus this $\sigma_8$ should be regarded 
as a measure of the initial perturbations. 

Our analyses concern high peaks that are physically related to halos with $M\sim 10^{14}{M}_{\odot}$ and above.
The baryonic effects on their mass function and overall density profiles are shown to be minimal 
\citep[e.g.,][]{Sawala2013, Schaller2015a, Schaller2015b}. 
Depending on baryonic physics, the very central part of halos may be affected \citep[e.g.,][]{Schaller2015b}.
However, our smoothing operation can suppress effectively the influence of detailed central profiles.  
We therefore do not expect significant baryonic effects on high peak abundances for the current WL data with relatively 
large statistical errors \citep[e.g.,][]{Osato2015}.

We focus on deriving constraints on $(|f_{R0}|, \Omega_{\rm m}, \sigma_8)$, the parameters that WL effects are most sensitive to.

We employ priors on $\Omega_{\rm m}$ and the initial curvature perturbation parameter $A_{\rm s}$ 
from WMAP9 \citep{Hinshaw2013} or Planck15 \citep{Planck2015}, where $A_{\rm s}$ can be directly linked to $\sigma_8$ \citep[e.g.,][]{Schmidt2009}.
Thus our total $\chi^2$ is 
\begin{equation}
\chi_{\rm tot}^2=\chi_{\rm peak}^2+\chi_{\Omega_{\rm m}}^2+\chi_{A_{\rm s}}^2.
\label{chitot}
\end{equation}
Here $\chi_{\Omega_{\rm m}}^2=(\Omega_{\rm m}-\Omega_{\rm m}^{\rm prior})^2/\sigma_{\Omega_{\rm m}^{\rm prior}}^2$ and 
$\chi_{A_{\rm s}}^2=(A_{\rm s}-A_{\rm s}^{\rm prior})^2/\sigma_{A_{\rm s}^{\rm prior}}^2$, 
where $\Omega_{\rm m}^{\rm prior}$ and $A_{\rm s}^{\rm prior}$ 
are the prior central values, and $\sigma_{\Omega_{\rm m}^{\rm prior}}$ and $\sigma_{A_{\rm s}^{\rm prior}}$ 
are the corresponding 68\% confidence limits. The specific priors are listed in Table \ref{tab:prior}.
These priors do not include the contributions from
the constructed lensing potential that depends on gravity theories.
On the other hand, for the small $|f_{R0}|$ concerned here, the impacts of modified gravity on the primordial cosmic microwave background (CMB)
and on the late integrated Sachs-Wolfe (ISW) effect are negligible \citep[e.g.,][]{Song2007, Planck15XIV}.
Thus the priors we adopt here that are derived under $\Lambda$CDM are feasible and should not introduce biases to our
constraints on $|f_{R0}|$.
The other cosmological parameters, such as the baryon density $\Omega_{\rm b}$, the Hubble constant $h$ and 
the power index of initial density perturbations $n_{\rm s}$, are fixed to the
corresponding values of WMAP9 or Planck15.
 
\section{Mock tests}

To validate our pipeline, we generate mocks from ray-tracing simulations. 

We run N-body simulations for flat $\Lambda$CDM under GR, and for $f(R)$ theory with $n=1$
and $|f_{R0}|=10^{-4}$ (F4), $10^{-5}$ (F5) and $10^{-6}$ (F6), respectively. 
Besides $|f_{R0}|$, all the other cosmological parameters are the same in all simulations with
$\Omega_{\rm m}=0.281$, $\Omega_\Lambda=0.719$, $\Omega_{\rm b}=0.046$, $h=0.697$,
$n_{\rm s}=0.971$ and the $\Lambda$CDM-equivalent $\sigma_8=0.819$.

The simulations start at redshift $z=49$ with the initial conditions generated by MPgrafic \citep{Zhao2011a}.
The ECOSMOG \citep{Li2012c} is used for the dynamical evolutions. The box size is $1024h^{-1}\hbox{Mpc}$ and the particle number is $1024^3$.
{We compare the halo mass function from these simulations with the predictions from \citet{Kopp2013}, and find a good agreement.}

The mock WL analyses for GR, F5 and F4 are done. Here we mainly present the results for GR and F5.
For each model, we run five independent N-body
simulations and pad them together to form the light cones to $z=3$. The five simulations
for F5 have exactly the same initial conditions as their GR counterparts.

Based on the padded simulations, we then use $36$ lens planes evenly distributed in the comoving distance to $z=3$ to 
perform multiple-plane ray-tracing calculations following closely the procedures applied in our previous studies\citep{Liu2014, Liu2015}. 
To generate mock data, we divide the simulated area into different fields of $1\time 1\hbox{ deg}^2$, and 
match them randomly to the observational fields. In each field, we preserve the relative positions and the photo-zs of the
observed galaxies, as well as the masked areas.  
We randomly rotate source galaxies to eliminate the original WL signals, and then incorporate 
the reduced WL shears from ray-tracing simulations to construct the mock shear data.
To better estimate the shape noise effects, we apply $15$ sets of different random rotations to the source galaxies. Thus for each model, 
we finally have $15$ sets of mock data, each with a survey area of $\sim ~150\hbox{ deg}^2$. 
We refer the readers to \citet{Liu2015} for further details. 

With mock data, we perform the same WL peak analyses as we do for observational data, and derive cosmological constraints
to validate our analyzing pipeline.   

\begin{table}
\caption{Summary of the prior information for different cases.} 
\label{tab:prior}
\begin{center}
  \leavevmode
    \begin{tabular}{C{1.6cm}C{2.5cm}C{2.2cm}C{2cm}}\hline
               Parameter  &  Obs. (WMAP9) (WMAP+BAO+$H_0$)& Obs. (Planck15) (TT,TE,EE+LowP) &  Mock  (F5\&GR)\\
               \hline
               $\Omega_{\rm m}^{\rm prior}$  &  $0.288\pm 0.0093$ & $0.3156\pm 0.0091$ & $0.281\pm 0.0093$\\
               $10^9 A_{\rm s}^{\rm prior}$   &       $2.427\pm 0.079$ & $2.207\pm 0.074$ & $2.372\pm 0.079$\\
               $k_{\rm pivot} ({\rm Mpc}^{-1})$ & $0.002$ & $0.05$ & $0.002$\\
               $\Omega_{\rm b}$ & $0.0472$ & $0.0492$ & $0.046$\\
               $h$ & $0.6933$ & $0.6727$ & $0.697$\\
               $n_{\rm s}$ & $0.971$ & $0.9645$ & $0.971$\\
               \hline
     \end{tabular}
    \end{center}
\end{table}

\begin{figure}
\includegraphics[width=0.495\columnwidth, height=0.36\columnwidth]{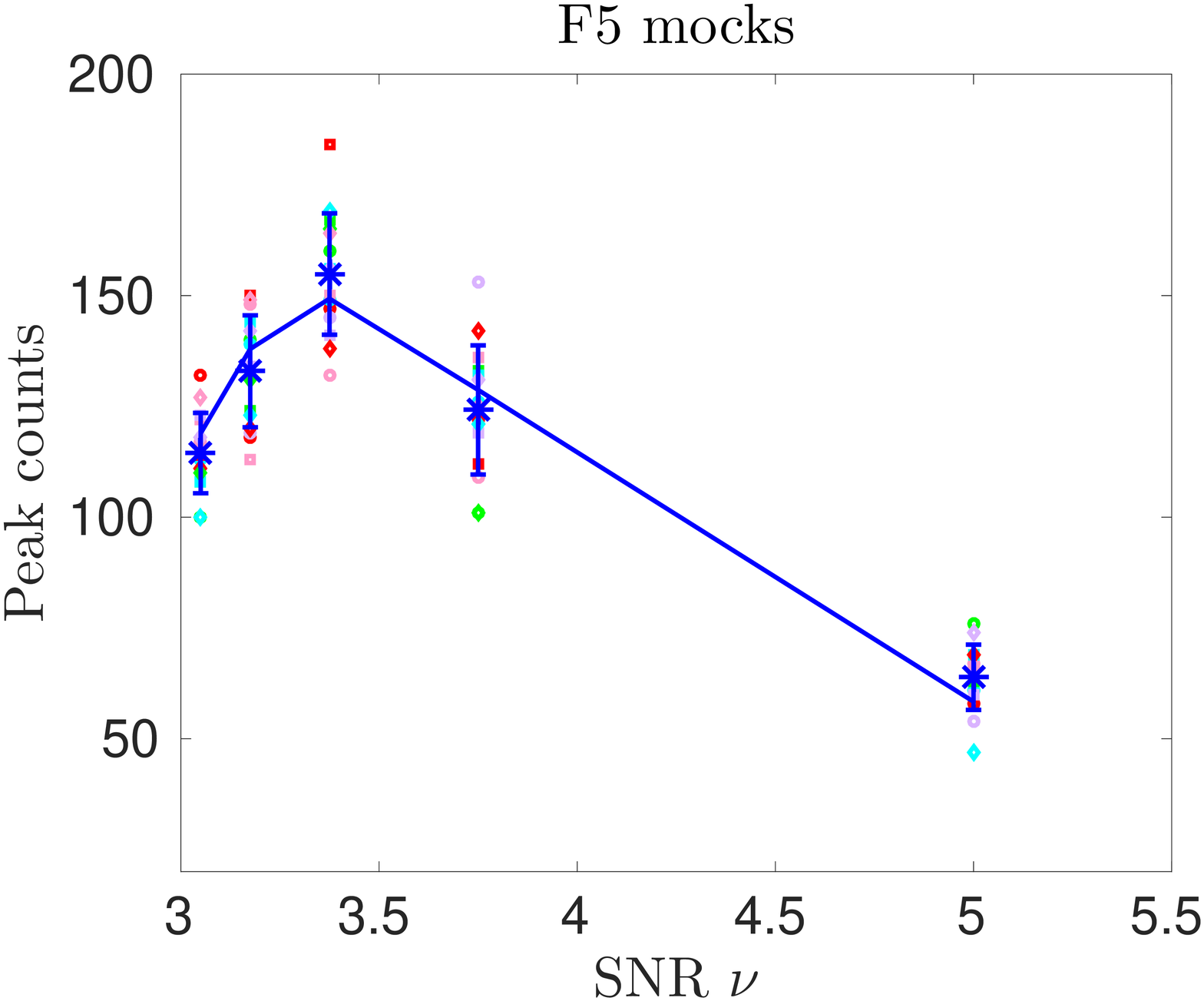}
\includegraphics[width=0.495\columnwidth, height=0.36\columnwidth]{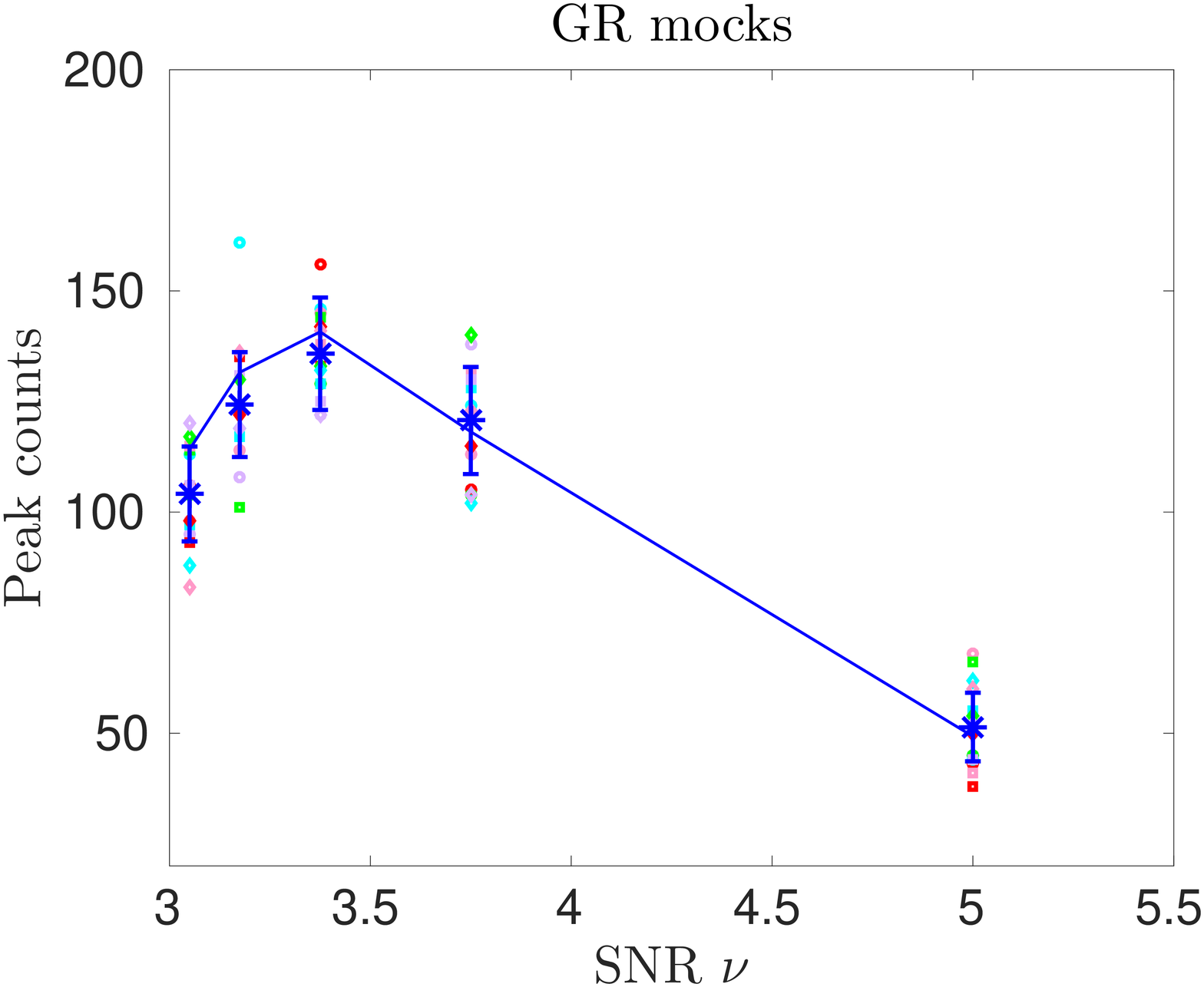}
\includegraphics[width=0.9\columnwidth, height=0.32\columnwidth]{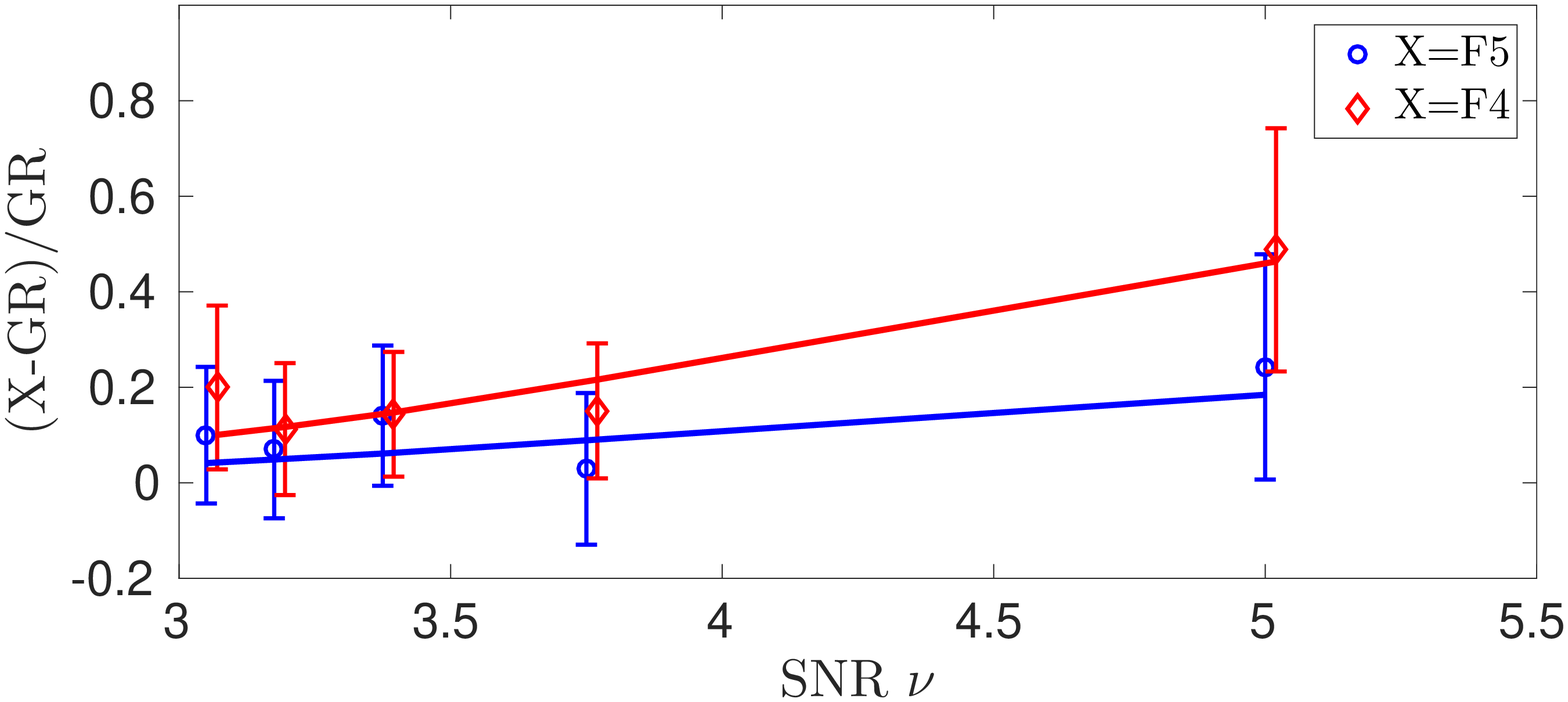}
\caption{\label{fig:mockpeak} Peak counts distribution from F5 (upper left) and GR (upper right) mock simulations. 
Different symbols with different colors correspond to different noise realizations. The blue `*' and the error bars are for 
the average values and the rms over the 15 realizations. The solid line is for our model predictions.
The lower panel is for the difference ratios.}
\end{figure}

\section{Results}

We first present the results from mock simulations.
Figure~\ref{fig:mockpeak} shows the peak number distributions for F5 (upper left) and GR (upper right). 
In both cases, the averaged mock results agree with our model predictions very well. 
The lower panel shows the difference ratios between F5 and GR (blue) and F4 and GR (red), respectively, 
which demonstrates the constraining potential of WL peak statistics on $|f_{R0}|$.

\begin{figure}
\includegraphics[width=1\columnwidth, height=0.5\columnwidth]{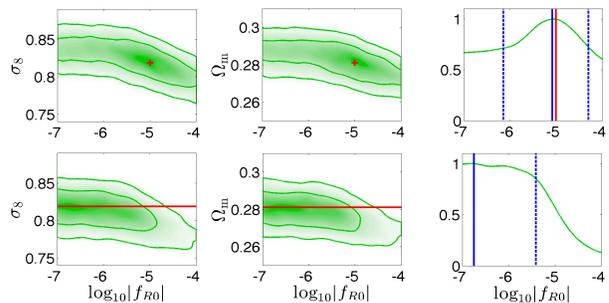}
\caption{\label{fig:mockfitting} Constraints derived from F5 (upper) and GR (lower) mock data. 
In 1-d distributions, blue solid and dashed lines indicate the locations of the maximum marginalized probabilities, 
and the corresponding 68\% confidence intervals. Red solid lines and `+' symbols are the input parameters of the mock simulations.
For the GR case, $f_{R0}=0$, and we only indicate the input $\Omega_{\rm m}$ and $\sigma_8$. }
\end{figure}

With these averaged data as our mock `observed' data and the covariance matrix derived by bootstrapping from the $15$ sets of 
simulated catalogs,
we perform Markov Chain Monte Carlo (MCMC) constraints on ($|f_{R0}|, \Omega_{\rm m}, \sigma_8$) using COSMOMC \citep{Cosmomc2002} modified to include our 
likelihood function from WL peak abundances. As in real observational analyses, we also add the priors on $\Omega_{\rm m}$ and $A_{\rm s}$
(Table \ref{tab:prior}). Here the central value of $\Omega_{\rm m}$ is directly from the simulation input, and
the $A_{\rm s}$ value is chosen to match the input $\sigma_8$. The $1\sigma$ ranges for the two parameters are taken from WMAP9. 
For $|f_{R0}|$, because its value spans orders of magnitude, we sample it in log-space
and apply a flat prior in the range of $\log_{10}|f_{R0}|=[-7, -4]$.

The obtained constraints are shown in Figure \ref{fig:mockfitting} for F5 (upper) and GR (lower).
The red symbols and lines denote the input values of the corresponding mock simulations.
It is seen that the 1-d maximum probability values (blue solid lines) agree with the input parameters excellently. 
Similar results are also obtained for F4 mock analyses.
The flattening trend for $\log_{10}|f_{R0}|< {-6}$ in the GR case is a reflection 
of the non-detectable differences for high peak abundances between GR and $f(R)$ with $|f_{R0}|< 10^{-6}$ due to the
chameleon effect. 
The 1-d marginalized constraints on $\log_{10}|f_{R0}|$ for GR and F5 mocks are shown in Table \ref{tab:constraint}. 
For F5, we show the $68\%$C.L. because the $95\%$C.L. is beyond our considered ranges of $\log_{10}|f_{R0}|$.

We now show the observational results from CFHTLenS.
We note that in the base Planck15 constraints, a minimum neutrino mass of $0.06\hbox{ eV}$ is included in their analyses. To be consistent, 
we therefore also include this neutrino mass in our peak abundance calculations when the Planck15 priors are applied.   

\begin{figure}
\includegraphics[width=1\columnwidth, height=0.7\columnwidth]{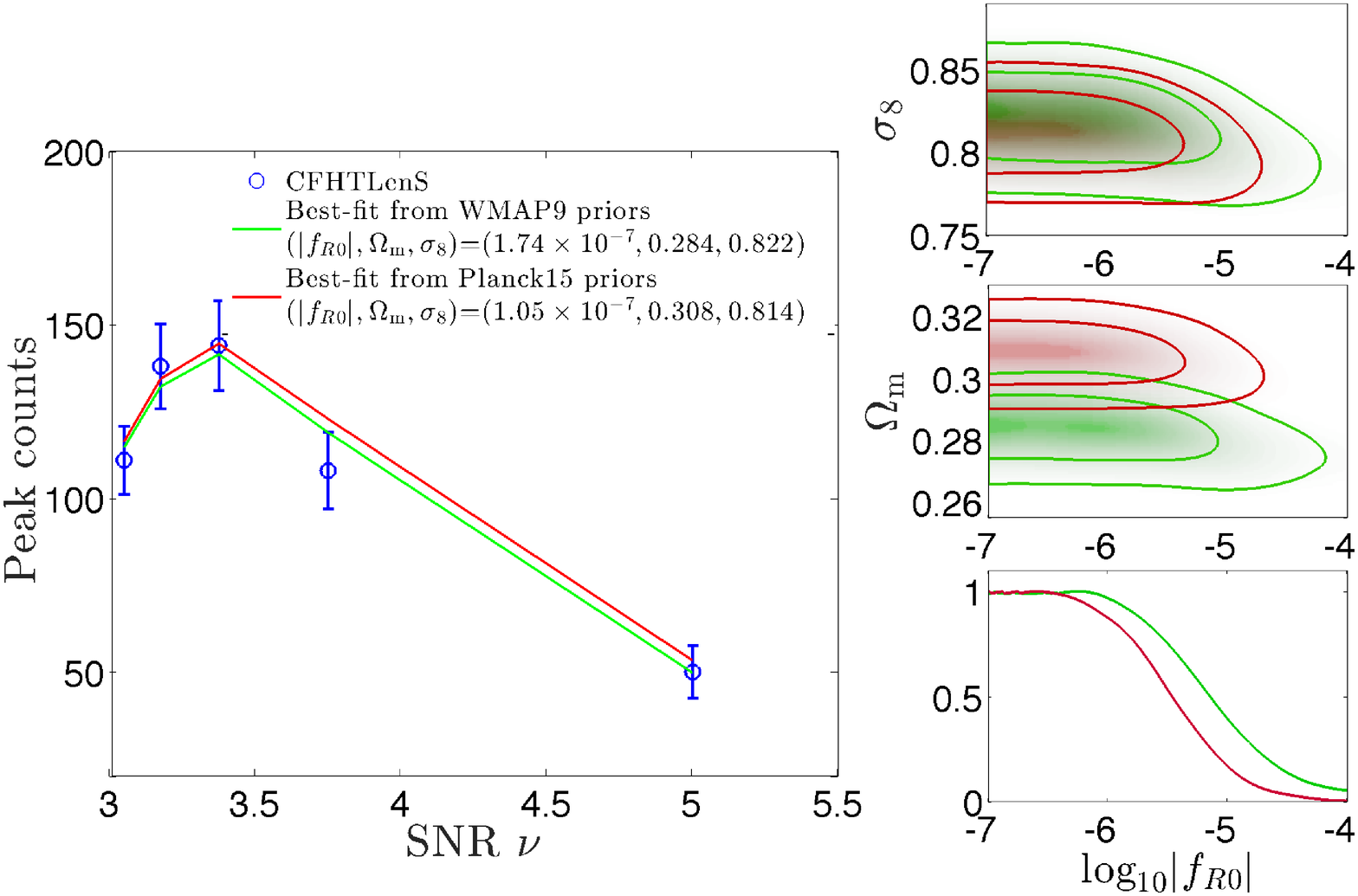}
\caption{\label{fig:obs_results} 
Results from CFHTLenS observational data. Left: The peak counts distribution. The corresponding solid lines are 
the theoretical predictions with the best-fit cosmological parameters listed therein.
The error bars are the square root of the diagonal terms of the covariance matrix. Right: The derived constraints.
Green and red contours are the results with WMAP9 and Planck15 priors, respectively.}
\end{figure}

The results are presented in Figure \ref{fig:obs_results}. The left panel shows the peak count distribution 
along with the theoretical predictions from the best-fit cosmological parameters obtained from the MCMC fittings using WMAP9 (green) 
and Planck15 (red) priors, respectively. The right panels show the derived constraints.
The marginalized 1-d constraints for $|f_{R0}|$ are shown in Table \ref{tab:constraint}, 
where the results from linear sampling on $|f_{R0}|$ and the value whose posterior probability is $\exp(-2)$ ($2\sigma$)
of the maximum probability are also listed.  

It is seen that WL peak abundance analyses can provide strong constraints on $|f_{R0}|$ even 
with data from surveys of an area of $\sim 150\deg^2$. 
The $95\%$C.L. from log-space sampling is $\log_{10}|f_{R0}|<-4.82$ and $<-5.16$ with WMAP9 and Planck15 priors, respectively.
The stronger constraint from Planck15 is due to its somewhat larger value of $\Omega_{\rm m}$.

Our constraints are comparable and slightly tighter than that from \citet{Cataneo2015} with 
$\log_{10}|f_{R0}|<-4.73$ (WMAP9) and $<-4.79$ (Planck2013) noting their wider prior of $[-10, -2.523]$ on $\log_{10}|f_{R0}|$. 
Comparing to the results in Table 8 of \citet{Planck15XIV}, our equivalent constraint on $B_0$ is $B_0<2.45\times 10^{-4}$ (Planck15)
(linear sampling), which is about $2-3$ times larger than theirs obtained by adding data of redshift space distortion 
and WL 2-pt correlations to Planck CMB data.

In the analyses above, we fix ($n_s$, $h$, $\Omega_{\rm b}$). WL peak abundances
depend on them very weakly. On the other hand, given $A_s$, the derived $\sigma_8$ changes with their values,
which in turn may affect our constraint on $|f_{R0}|$. To test this, we perform MCMC analyses by 
including them separately as additional free parameters and applying WMAP9 priors, which are larger than those of 
Planck15. We find that by adding $n_s$ or $h$ or $\Omega_{\rm b}$, the constraint is 
weakened by $\sim 1.4\%,0.9\%$ and $ 0.2\%$ with $\log_{10}|f_{R0}|<-4.75$, $-4.78$, and $-4.81$, respectively. 
Considering the negative degeneracy between $n_s$ ($h$) and $A_s$ from WMAP9, their influences on 
the $|f_{R0}|$ constraint should be even smaller.

\section{Summary}

Using CFHTLenS, we derive constraints on $f(R)$ theory, for the first time using WL peak abundance analyses.
To demonstrate the potential of the probe, 
we focus on the specific HS07 model with $n=1$. We find no evidence of
deviations from GR and obtain strong limits on the $|f_{R0}|$ parameter. 
For other $n$ values with $n>1$, because of the $|f_{R0}|-n$ degeneracy \citep[e.g.,][]{Li2011},
we expect that the limit on $|f_{R0}|$ would be larger. We will perform more general studies in the future.

WL high peaks are closely associated with massive clusters, 
and thus the constraining power of WL high peak abundances is physically similar to that of cluster abundance studies.
However, WL peak analyses are much less affected by baryonic physics than other cluster probes in which
baryon-related observables are involved.
On the other hand, the WL peak signal depends on the halo density profile, 
whose shape is determined by the concentration parameter for a NFW halo. Thus the 
uncertainty in the halo M-c relation can potentially affect the cosmological constraints from WL peak abundances. 
This impact is weak for our current analyses given the data statistics. 
For future large observations, such effect needs to be considered carefully.
Our studies show that we can constrain the M-c relation simultaneously with cosmological parameters from WL peak counts
to avoid potential biases from the assumed M-c relation \citep{Liu2015}.

With improved WL data, we expect that our analyses can be applied to constrain a more general class of modified gravity theories that
can affect the halo abundances significantly.

\begin{table}
\caption{Constraints from mock and observational analyses.}
\label{tab:constraint}
\begin{center}
  \leavevmode
    \begin{tabular}{c c c c} \hline
                             &  Mock &  &\\
                             \hline
                Parameter &  case  &  &   \\
                                         \hline
               $\mathrm{log}_{10}|f_{R0}|$ \footnotemark[1]
                   & GR (1-d 95\% limit) & $<-4.59$  &\\
               $\mathrm{log}_{10}|f_{R0}|$ \footnotemark[1]
                   & F5 (1-d best fit and 68\%C.L.) & $-5.08^{+0.81}_{-1.06} $ & \\
    \hline
    \hline
                             &  CFHTLenS observation & & \\

     \hline
                Parameter &  case  &  WMAP9 &   Planck15 \\
                                         \hline
               $\mathrm{log}_{10}|f_{R0}|$ \footnotemark[1]   
                   & 1-d limit (95\%) & $<-4.82$ & $<-5.16$ \\
               $|f_{R0}|$ \footnotemark[2] 
                   & 1-d limit (95\%) & $<7.59\times10^{-5}$ & $<4.63\times10^{-5}$ \\
               $\mathrm{log}_{10}|f_{R0}|$ \footnotemark[3] 
                   & 1-d limit ($2\sigma$) & $<-4.50$ & $<-4.92$ \\
               \hline

     \end{tabular}
    \footnotetext[1]{Probability distribution obtained based on $\mathrm{log}_{10}|f_{R0}|$}
     \footnotetext[2]{Probability distribution obtained based on $|f_{R0}|$}
     \footnotetext[3]{$\exp(-2)$ of the maximum probability in log-space}
    \end{center}
\end{table}

\section{Acknowledgement}
This work used the DIRAC Data Centric system at Durham University, operated by the Institute for Computational Cosmology on behalf of the STFC DIRAC HPC Facility (www.dirac.ac.uk). This equipment was funded by BIS National E-infrastructure Capital Grant No. ST/K00042X/1, STFC Capital Grants No. ST/H008519/1 and No. ST/K00087X/1, STFC DIRAC Operations Grant No. ST/K003267/1 and Durham University. DIRAC is part of the National E-Infrastructure.
The MCMC calculations are partly done on the Laohu supercomputer
at the Center of Information and Computing at
National Astronomical Observatories, Chinese Academy of Sciences,
funded by Ministry of Finance of People's Republic of China
under Grant No. ZDYZ2008-2. 
The research is supported in part by NSFC of China under Grants No. 11333001, No. 11173001, and No. 11033005.
L.P.F. also acknowledges the support from NSFC Grants No. 11103012, Shanghai Research Grant No. 13JC1404400 $\&$ No. 16ZR1424800 of STCSM, and from Shanghai Normal University Grants No. DYL201603. G.B.Z. is supported by the 1000 Young Talents program in China.
Z.H.F. and G.B.Z. are also supported by the Strategic Priority Research Program “The Emergence of Cosmological Structures”
of the Chinese Academy of Sciences Grant No. XDB09000000. B.L. acknowledges the support by the U.K. STFC Consolidated Grant No. ST/L00075X/1 and No. RF040335. 
M.C.C. acknowledges the support from NSFC Grants No. 1143303 and No. 1143304. X. K. L. acknowledges the support from General Financial Grant from China Postdoctoral Science Foundation with Grant No. 2016M591006.
To access the simulation data used in this work, please contact the authors.

\bibliography{ms}

\end{document}